# Polaronic transport in the ferromagnetic phase of $Gd_{1-x}Ca_xBaCo_2O_{5.53}$

N. Thirumurugan, C. S. Sundar and A. Bharathi*

*Condensed Matter Physics Division, Materials Science Group, Indira Gandhi Centre for Atomic Research, Kalpakkam, India*

## Abstract

Temperature dependent electrical resistivity and thermopower measurements were carried out on $Gd_{1-x}Ca_xBaCo_2O_{5.53}$ with x varying between 0 and 0.25. Ca subsitution leads to the incorporation of holes ($Co^{4+}$) into the system that leads to a reduction in resistivity and a stabilisation of the ferromagnetic phase at low temperatures. The temperature dependence of resistivity and thermopower are markedly different in the Ca doped sample, with a dramatic reduction in the resistivity, as compared to that in the pristine sample. The variation in both the resistivity and thermopower with temperature is explained in terms of the transport of polarons in the ferromagnetic phase of Ca doped system.

*Corresponding Author
A. Bharathi
Head, Low Temperature Studies Section
Condensed Matter Physics Division
Materials Science Group
IGCAR, Kalpakkam, 603102
Phone No.27480500-22356
E-mail: bharathi@igcar.gov.in

## 1. Introduction

The oxygen deficient double perovskite $GdBaCo_2O_{5.5\pm\delta}$ has been subjected to much study in the last few years[1-5]. $GdBaCo_2O_{5.5}$ is seen to undergo several transitions viz., metal to insulator, para to ferromagnetic and ferro to anti-ferromagnetic with decrease in temperature. Like in other transition metal oxides, competing interactions viz., crystal field effects, Hunds rule coupling, and Coulomb interactions make a variety of ground states possible in this system. Further, the structure permits oxygen content variation anywhere between 5 and 6, with a systematic occupation of the O in the Gd-O layer in $GdBaCo_2O_{5.5\pm\delta}$. This results in Co valence changes from 1:1 mixtures of $Co^{2+}$ and $Co^{3+}$ and $Co^{3+}$ and $Co^{4+}$ at the extreme oxygen compositions. At the O concentration of 5.5, all Co ions are in the 3+ state. For O less than 5.5 the system is electron doped, whereas for O in excess of 5.5 it is hole doped. The transport behavior shows a strong asymmetry to hole and electron substitution with the former having a poor metallic ground state and the latter an insulating ground state[2]. Apart from tuning of properties by oxygen content variations, there have been several studies on chemical substitutions at the Co and the Gd site [6-10] in $GdBaCo_2O_{5.5}$, with a view to explore the role of electron and hole doping on the transport and magnetic properties.

For example, in Ni[6] substitution at the Co site, that leads to electron doping, the temperature dependence of resistivity shows a feature corresponding to the PMM to PMI transition, and also a weak signature corrosponding to the FMI to AFI transition [2]. The magnetic phase diagram has been mapped using magnetisation and magnetoresistance measurements. Sr substitution at Gd site [9] resulted in a strong decrease in the resistivity along with the stabilization of the ferromagnetic state. However, complete metallisation could not be realised, possibly because the maximum solubility of Sr at Gd site was restricted to ~7at% [9]. Since the ionic radius of Ca is smaller than that of Sr, it can be expected that the Ca substitution can metallise $GdBaCo_2O_{5.5}$. In the present study, we have explored the role of Ca doping on structure, transport and magnetic properties. We have synthesised $Gd_{1-x}Ca_xBaCo_2O_{5.53}$ for various Ca fractions, $0\leq x\leq 0.25$. Through an analysis of the temperature dependence of resistivity and thermpower, we argue that the system exhibits a polaronic transport – that is correlated with the ferromagnetic state.

## 2. Experimental Details

The $Gd_{1-x}Ca_xBaCo_2O_{5.53}$ samples with x upto 0.25, were prepared by solid state reaction from high purity, $Gd_2O_3$, $BaCO_3$, $CaCO_3$ and $Co_3O_4$ as described in earlier studies [6-9]. The stoichiometric mixtures were decarbonated at 850 $^0$C followed by grinding, pelletising and sintering at 1050 $^0$C for 30 hours. The whole process was repeated at least three times to ensure homogeneity. The phase purity was checked by XRD measurements using a STOE diffractometer. Lattice parameters were inferred from the XRD data using the STOE program. The $Co^{4+}$ fraction introduced by Ca doping was determined by iodometric titration [10]. AC susceptibility measurements were carried out to investigate the magnetic transitions using a home built dipper cryostat set up, operating at 941 Hz. The resistivity measurements were carried out using the four-probe geometry using a dipper cryostat. Thermopower measurements were carried out in a home built exchange gas set up [11] using a dc technique.

## 3. Results and discussions

The lattice parameter variation with Ca content is shown in Fig.1. It is seen from Fig.1a that with the increase of Ca content, the a- lattice parameter decreases, the b-parameter increases, while the c-lattice parameter does not vary, and these changes are such that the overall cell volume decreases marginally. The decrease in the a-lattice parameter and the increase in b-lattice parameter results in a sizeable increase in the orthorhombicity, (b/2-a)/(b/2+a), as shown in Fig.1b. It is seen that the orthorhombicity increases with Ca content, that can be fitted to two line segments with a change of slope at x = 0.1. The $Co^{4+}$ concentration in excess of that seen in the pristine sample, as estimated by iodometry is shown in Fig.1c for various Ca fractions. The variation in $Co^{4+}$ content occurs with two distinct slopes for $x<0.1$ and $x>0.1$. Initially, upto x= 0.1, the increase in $Co^{4+}$ content, follows the Ca concentration, beyond which the slope decreases. The oxygen stoichiometry estimated from charge balance is also shown in Fig.1c. A marginal decrease in O content is seen from 5.53 in the pristine sample to 5.47 in the x=0.20 sample. The rate of decrease in O content also shows distinct change at a Ca fraction of x=0.1. Despite the observed decrease in O content, a clear increase in $Co^{4+}$ is observed in Fig.1c, suggesting that Ca substitution is responsible for hole doping $GdBaCo_2O_{5.53}$. The depletion of oxygen is possibly responsible for the change in orthorhombicity as seen in Fig. 1b. Fig.2 shows the results of ac susceptibilty measurements done on doped samples. The sharp peak in $\chi'$ is associated with the paramagnetic to ferromagnetic transition [12] and the $T_C$ shows a systematic increase with the Ca content. In the case of the undoped sample, the weak signature corresponding to the FM to AFM transition [12] can be discerned. For x greater than 0.05, the transition to the AFM state is suppressed, and the sample remains in the ferromagnetic state at lower temperatures. The inset to Fig.2 shows the variation of $T_C$ with Ca content, x, and this again shows two distinct slopes, similar to the case of orthorhmbicity and $Co^{4+}$ content variations, shown in Fig.1. These correlated changes observed in Co valence, orthorhombicity and $T_C$ show the intimate coupling between the spin, lattice and charge coupling respectively, in this system.

In the following, we explore the effect of Ca doping on the transport properties, viz. resitivity and thermopower in samples with Ca fraction $0 \leq x \leq 0.1$, wherein the changes in $Co^{4+}$ content occurs mainly due to Ca doping, and not due to change in oxygen stoichiometry ( See Fig.1). Fig.3a shows the variation in resistivity of samples with Ca fractions x= 0 and 0.03. The pristine sample shows a rapid increase in resistivity with decrease in temperature, whereas it shows a less marked increase in the x=0.03 sample. For the x = 0.05 and x= 0.1 samples, an interesting feature of the resistivity behaviour is the presence of bump in the 150 K to 250 K range, as shown in Figs.3b and 3c respectively. Another way to illustrate the different temperature dependencies for various Ca contents, is to plot the resistivity values at three representative temperatures, viz., 10 K, 50 K, and 290K, as shown in the inset of Fig.3c. It is seen that with the incorporation of Ca, the resistivity at room temperature is seen to decrease, and there is a dramatic reduction in resistivity by over seven orders of magnitude at 10 K, for x= 0.1 sample when compared to that in the pristine sample. In order to understand the temperature dependence of resistivity, we first analyse the data in terms of variable range hopping(VRH) model for conductivity, viz., $\rho=\rho_0(\exp(T_o/T)^{1/4})$ [2]. This is shown in Fig. 4a. It is seen that while in the pristine sample, the VRH model well describes the temperature dependence of resistivity variation over the entire temperature range of 5 K to 300 K, starting from x=0.03 sample, there is a marked deviation from the VRH behavior, for the Ca substituted samples.

The hump-like feature seen in the temperature dependence of resistivity in the x=0.05 and x=0.1 samples, shown in Fig. 3b and 3c is reminiscent of polaronic transport – as seen for example in $TiO_2$ [13]. It is well known that a transition in the nature of carriers from bound polarons at low temperatures to free carriers at elevated tempeartures leads to a characteristic resitivity variation, given by $\rho = \rho_0 T \exp(E_\rho/k_B T)$[14], where $E_\rho$ is the polaron activation energy. With the above model in mind, in Fig. 4b, we plot ln(ρ/T) versus 1/T for samples with Ca fractions, x=0.05 and 0.1. In the temperature range of 50 to 200 K, the above model appears to give a reasonable fit. However, for the x=0.03 sample, (see Fig. 5a), it is seen that the polaronic model does not give an adequate description, viz., ln(ρ/T) is not linear with 1/T . In particular, a distinct change of slope is seen around 140 K, that coincides with the ferromganetic to anti ferromagnetic transition at low temperature [2-6], as seen from a plot of the magnetisation vs tempertaure shown in Fig. 5b. This figure also shows that with the application of a magnetic field of 6T, this transition to the anti-ferromagnetic phase is suppressed. Interestingly, when the transition to the antiferromagnetic phase is suppressed, the temperature dependence of resistivity is well described by the polaronic model, as can be seen from the linear variation seen in Fig. 5c, in the presence of magnetic field of 6 Tesla. Thus, Figs. 4 and 5c clearly point out that the temperature dependence of resistivity that is well described by the polaronic model, is linked to the stabilisation of the ferromagnetic phase. We now speculate on the nature of polarons in the ferromagnetic phase, taking cue from hole doped $La_{1-x}Sr_xCoO_3$[15]. Ca substitution results in the generation of low spin (LS)$Co^{4+}$, sites, that are also associated with structural distortion (as inferred from the orthorhombic distortion). This favours the surrounding $Co^{3+}$ octahedra, to tranform to an intermediate spin (IS), due to an altered crystal field. This configuration of LS $Co^{4+}$ and IS $Co^{3+}$ prefers a ferromagnetic interaction due to double exchange [15]. The hopping of LS $Co^{4+}$ into neighbouring IS $Co^{3+}$ moves along with it the distortion, thereby creating ferromagnetic polarons, in the entire volume of the sample.

The evidence for polaronic transport in the ferromagnetic phase is also seen in the thermopower measurements. It is known [2] that the $GdBaCo_2O_{5.5}$ system is characterised by a large thermopower that arises due to the entropy of transport that includes contributions from both the spin degeneracy and configurational degrees of freedom, as the charge carriers move, viz, the hopping of LS $Co^{4+}$ into an IS $Co^{3+}$ [16,17]. With two types of carriers, viz. free holes (h) and polarons (p) , the thermopower (S) can be viewed as a conductivity weighted average : $S=(\sigma_h S_h + \sigma_p S_p)/(\sigma_h + \sigma_p)$, as has been done in a related compound $La_{2-x}Sr_xCoO_4$ [14]. The results of thermopower measurements in the undoped and Ca doped samples is shown in Fig. 6. The results for the case of the undoped sample ( cf. Fig. 6a) is consistent with earlier measurements [2], and in particular shows the feature corresponding to the FM to AFM transition. In the $GdBaCo_2O_{5.5\pm\delta}$ system S at a given temperature measured as a function of $\delta$ shows a peculiar variation. The magnitude of S is zero for $\delta=0$; and increases steeply for small increase/decrease in $\delta$ and then decreases for larger $\delta$. This has been explained quantitatively in terms of configurational entropy of charge carriers enhanced by their spin and orbital degeneracies[18]. Likewise, in the present case too the thermopower at room temperature increases due to the addition of holes by Ca doping for x=0.03 as compared to the pristine sample and then decreases for higher Ca concentrations. For each Ca concentration, with the decrease of temperature, the thermopower shows a broad maximum followed by a rapid decrease. The change in sign of dS/dT at 250 K coincides with the change in the temperature coefficient of resistivity (see Fig.3b-c). As indicated earlier, in the temperature range below 250 K, the temperature dependence of resitivity was explained in terms of polaronic transport (see Fig.4).

It is known that for polaronic transport the temperature dependent thermopower S is given by $(k_B/e)(\alpha+E_S/k_BT)$ [14] where $k_B$ and e are the Boltzmann constant and electronic charge respectively and $\alpha$ depends on the nature of polarons and $E_S$ the activation energy for polaron hopping [14]. Fig.6b shows a linear variation in the plot of S versus 1/T in this temperature range, suggesting a polaronic transport. It is seen from Figs 6b that best fits are obtained with two component fits with distinct slopes at the high and low temperature regimes, viz, 180K-125K (HT) and 125K-77K (LT). The activation energies and values of $\alpha$ obtained from these fits are given in Table.1. The parameter $\alpha$ > 1suggests that the polarons are large. This is also seen in Fig. 4, and the results of the fit of temperature dependence of resistivity to polaronic model are indicated in Table.2. Given that the polaronic transport is linked with the ferromagnetic phase, the occurence of two slopes suggests the existence of two co-existing ferromagnetic phases. Magnetisation measurements give evidence for the co-existence of a ferromagnetic insulating and ferromagnetic metallic phase with different co-ercivities. The metallic FM phase at low temperatures (below 200 K) is stabilised by the double exchange mechanism and is associated with polaron hopping. The magnetic phase diagram of Ca doped $GdBaCo_2O_{5.53}$, using results of magnetisation hystersis loop and magnetoresistance measurements has been evaluated [19].

## 4.Summary

The resitivity and thermopower measurements in the Ca doped samples can be explained in terms of polaronic transport in the ferromagnetic phase. This is particularly evident from the behaviour seen in the x=0.03 sample, where the observation of polaronic transport is associated with the stabilization of ferromagnetism under applied magnetic field of 6 Tesla. The temperature dependence both the resistivity and thermpower are best fitted to two slopes. This is associated with the co-existence of two ferromagnetic phases, the evidence for the same is obtained from magnetisation and hysterisis loop measurements[19].

## Acknowledgements

The authors thank S. Paulraj, Periyar University, Salem, A. T. Satya, Condensed Matter Physics Division, IGCAR and S. Sen, Radiological Safety Division, IGCAR for their early involvement in the experiments.

## References

[1] I.O. Troyanchuk, N.V. Kasper, D.D. Khalyavin, H. Szymczak, R. Szymczak and M.Baran, Phys. Rev. Lett. **80**, 3380 (1998).

[2] A.A.Taskin, A. N. Lavrov, and Yoichi Ando**,** Phys.Rev.B **71**, 134414 (2005)

[3] A.A.Taskin and Yoichi Ando, Phys.Rev.Lett. **95,** 176603 (2005)

[4] A.Maignan, V. Caignaert, B. Raveau, D. Khomskii, and G. Sawatzky, Phys. Rev. Lett. **93** , 026401 (2004).

[5] D. D. Khalyavin, D.N. Argyriou, U. Amman, A. A. Yarenchenko and V. V. Kharton, Phys. Rev. B **75**, 134407 (2007)

[6] A. Bharathi, P.Yasodha, N. Gayathri, A.T.Satya, R.Nagendran, N.Thirumurugan, C.S. Sundar and Y. Hariharan Phys. Rev. B, **77**, 085113 (2008)

[7] N.Thirumurugan, A. Bharathi and C.S. Sundar, Journal of Magnetism and Magnetic Materials **322,** 152 (2010)

[8] P. Yasodha, N. Gayathri, A. Bharathi, M. Premila, C. S. Sundar and Y. Hariharan, Solid State Communications **144** 215 (2007**)**

[9] J. Janaki, A.Bharathi, N.Gayathri, P.Yasodha, M.Premila, V.S. Sastry, and Y. Hariharan, Physica B **403**,631 (2008)

[10] K. Conder, E. Pomjakushina, A. Soldatov and E. Miltberg, Materials Research Bulletin, **40**, 257 (2005)

[11] N. Gayathri, A. Bharathi, Y. Hariharan, Sol. State Phys. (India) 50 (2005) 699.

[12] A. Maignan, V. Caignaert, B. Raveau, D. Khomskii, and G. Sawatzky, Phys. Rev. Lett., 93, 026401 (2004)

[13] N. F. Mott and E. A. Davis, “*Electronic Processes in non-crystalline materials*”, Oxford University press.

[14]C.H.Sun, H.S.Yang, Y.S.Chai, J.liu, H.X.Gao, L.Cheng, J.B.wang and L.Z.Cao, J. Phys. Chem. Sol. **70** 286 (2009)

[15]D. Louca and J. L. Sarrao, Phys. Rev. Lett **91**, 15501 (2003)

[16] W. Koshibae, T. Tsutsui and S.Maekawa, Phys. Rev. B **62,** 6869 (2000)

[17]C. Frontera and Garcia-Munoz, EPL **84**, 67011 (2008)

[18] A. A. Taskin, A. N. Lavrov, and Yoichi Ando, Phys. Rev. B **73**, 121101(R) (2006)
[19] N. Thirumurugan, A. Bharathi and C. S .Sundar (To be published)

**Figure Caption**

**Fig.1 (a)** The variation of lattice parameters 'a' and 'b/2' in left panel and 'c/2' in righ panel as a function of Ca content. **(b)** The variation of orthorhombicity defined as (b/2-a)/(b/2+a) with Ca content. **(c)** Variation of $Co^{4+}$ content with respect to that in the pristine sample, as obtained by iodometric titration as a function of Ca fraction, the corresponding oxygen stoichiometry is also shown. Straight lines in the figures **(b & c)** are linear fits at different Ca fraction regimes $0 \leq x \leq 0.1$ and $x \geq 0.1$.

**Fig.2** The variation of ac susceptibility for representative Ca fractions shown in different panels. The temperature at which the peak appears identified as the Curie temperature $T_C$, whose varaition with Ca fraction is shown in the inset.

**Fig.3 (a)** Temperature dependent resistivity for Ca fractions x=0 and x=0.03. **(b)&(c).** Resistivity of samples x=0.05 and 0.1 respectively. Inset in **(c)** shows the change in resisitivity as a function of Ca content at temperatures indicated.

**Fig.4 (a)** VRH behavior of x=0, 0.03, 0.05 and 0.1 samples. Straight lines are guide to eye. **(b)** Variation of ln(ρ/T) versus 1/T for sample with Ca content x=0.05 and x=0.1 Linear fits to polaronic model are represented by lines.

**Fig.5 (a)** Variation of ln(ρ/T) versus 1/T for sample with Ca content x=0.03 **(b)** Field cool magnetization measurements for samples x=0.03 at H=0.01 Tesla and at H=6 Tesla **(c)** The variation of ln(ρ/T) versus 1/T in x=0.03 sample; ρ measured at 6 Tesla.

**Fig.6 (a)** Thermopower S versus 1000/T for Ca fractions x=0 and 0.03. Arrow indicates the anomaly at the ferromagnetic to anti-ferromagnetic transition ($T_N$) **(b)** S versus 1000/T for Ca fractions x=0.05 and x=0.1.

**Table.1** : The activation energies extracted from the slopes of the S versus 1/T plot at two temperature regimes (LT & HT) in meV; $\alpha$ are the corresponding constants extracted from thermopower fits.

| **Ca Fraction** | **$E_S$(LT) (meV)** | **$E_S$(HT) (meV)** | **α(LT)** | **α(HT)** |
|---|---|---|---|---|
| 0.05 | 19.6±0.3 | 25.16±0.3 | 3.8655 | 4.668 |
| 0.1 | 16.9±0.77 | 22.79±0.5 | 2.721 | 3.828 |

**Table.2 :** The activation energies extracted from the slopes of the ln ($\rho$/T) versus 1/T plot in the 77 K to 125 K temperature range (LT) and 125 K to 200 K temperature (HT) ranges in meV.

| **Ca Fraction** | **$E_\rho$(LT) (meV)** | **$E_\rho$(HT) (meV)** |
|---|---|---|
| 0.05 | 18.1±0.07 | 13.6±0.08 |
| 0.1 | 20.2±0.1 | 16.3±0.06 |
| 0.03<br>(at 6 Tesla) | 19.9±0.15 | 14.9±0.14 |

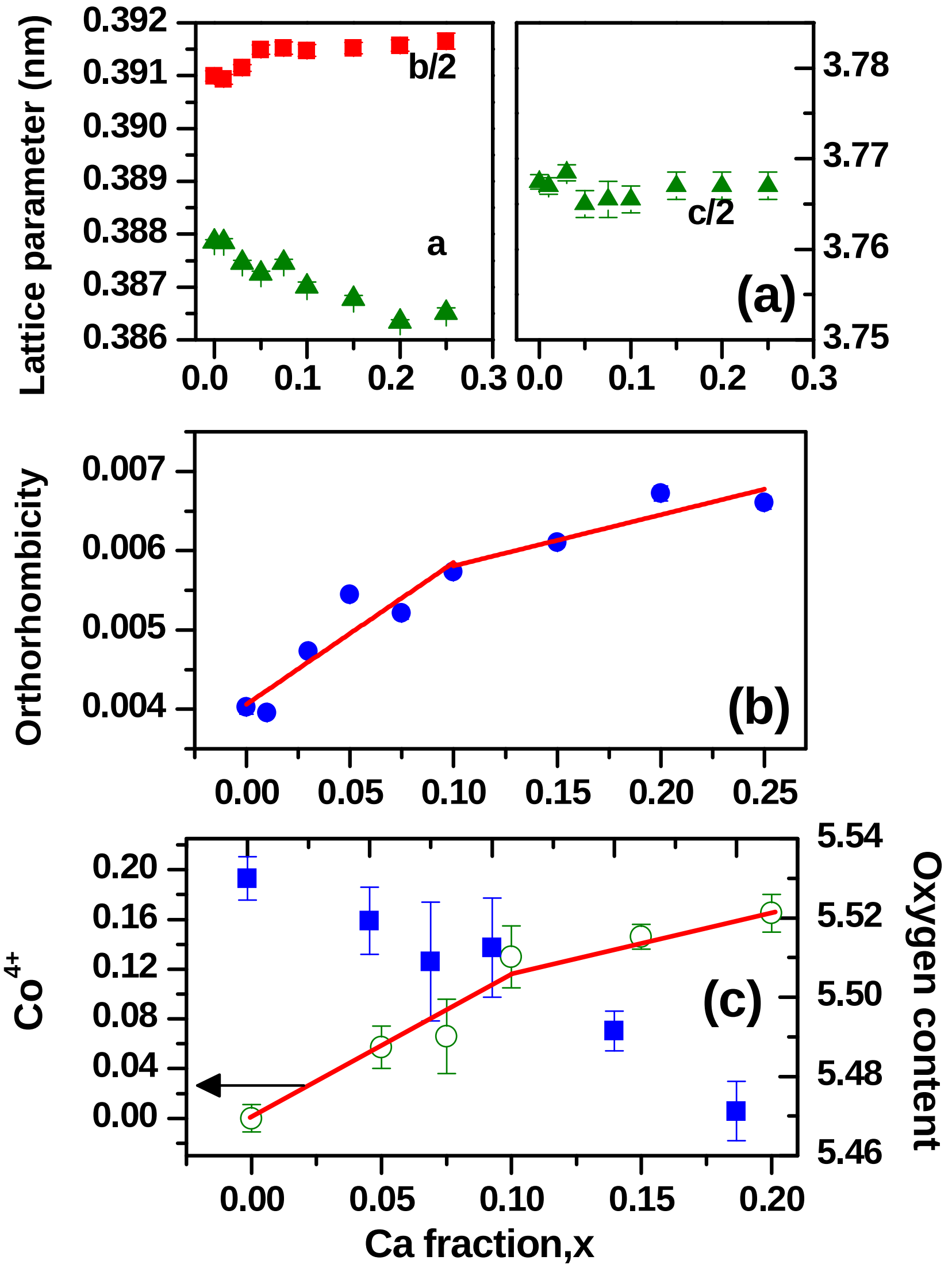

Figure 1

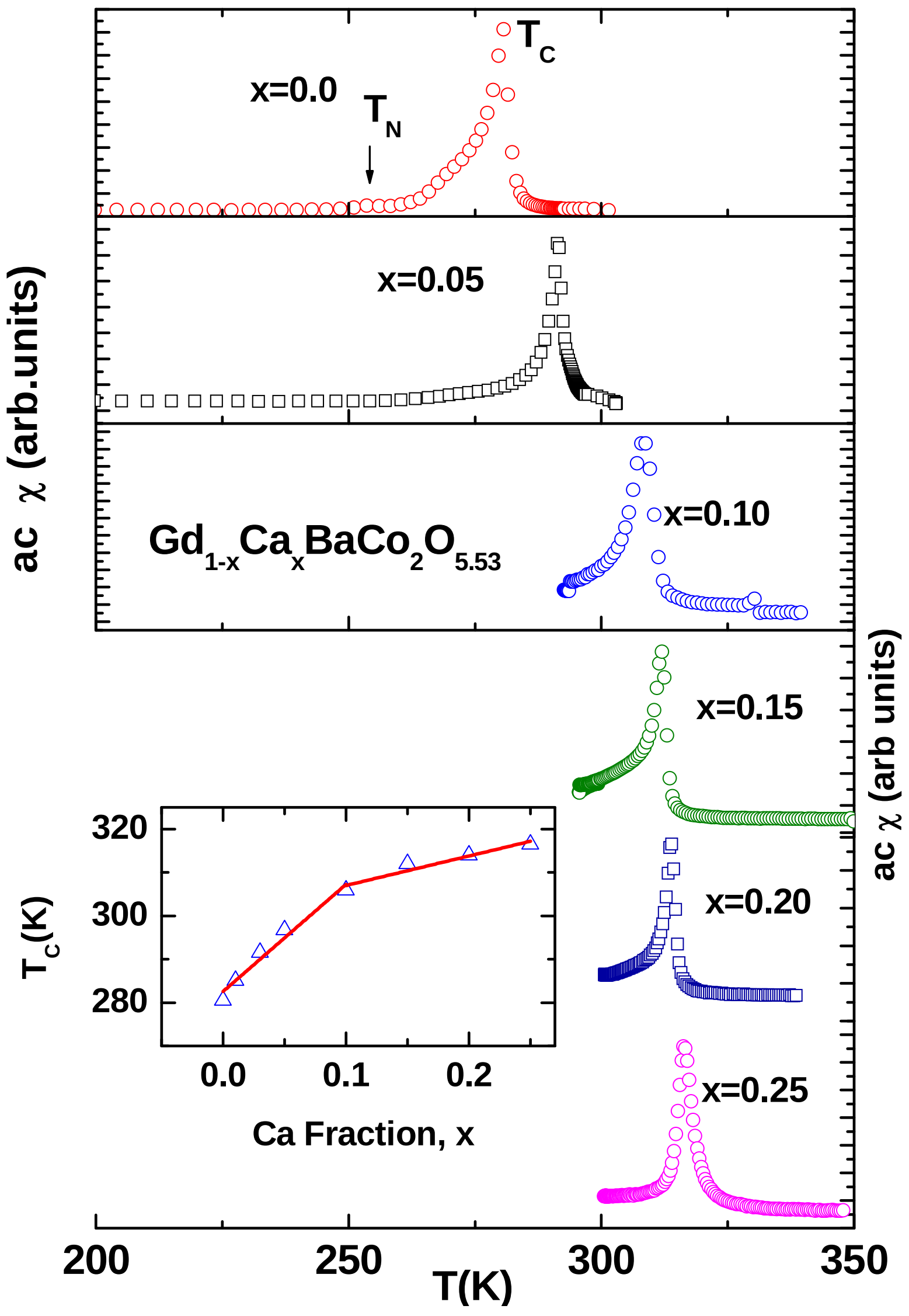

Figure 2

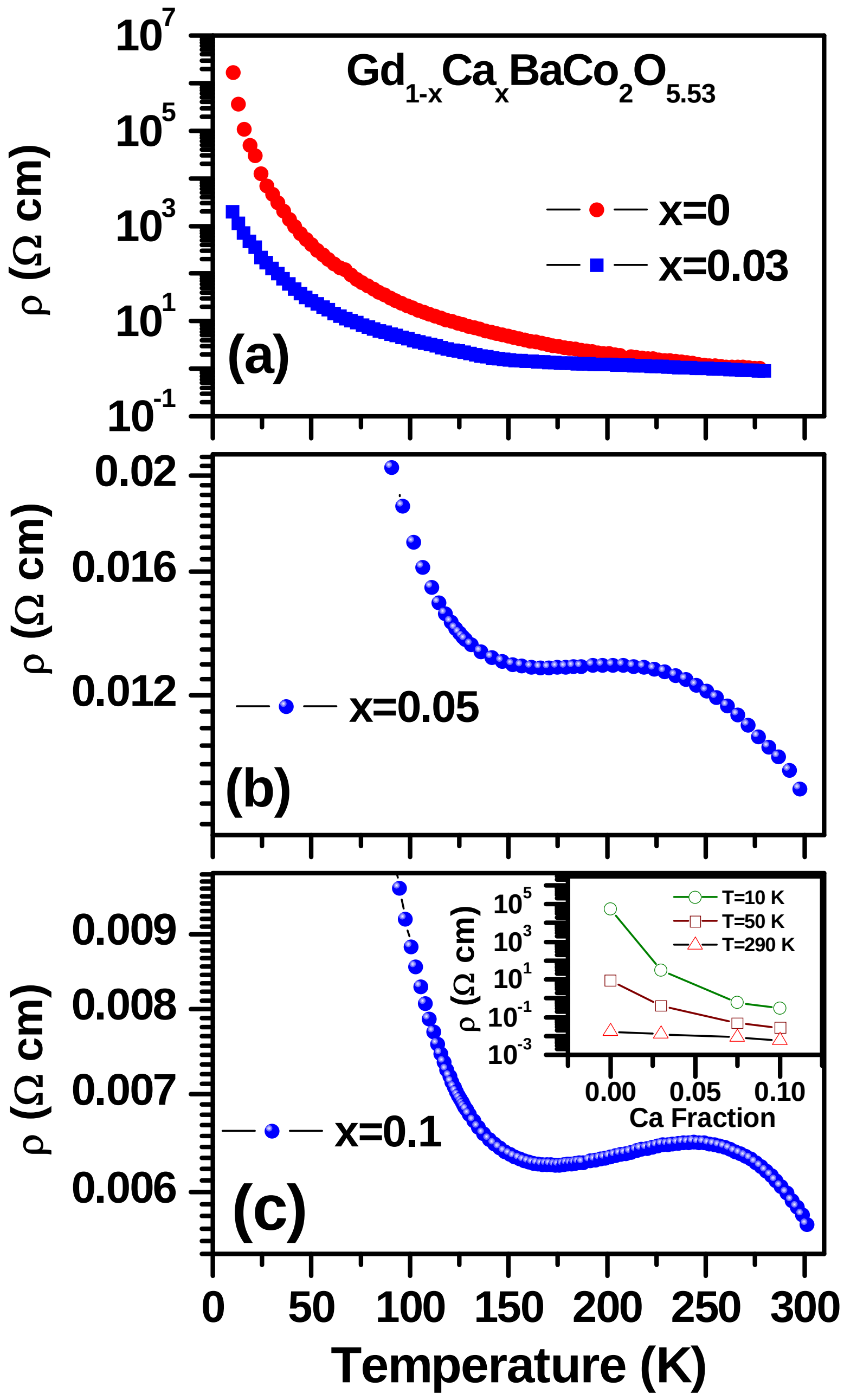

Figure 3

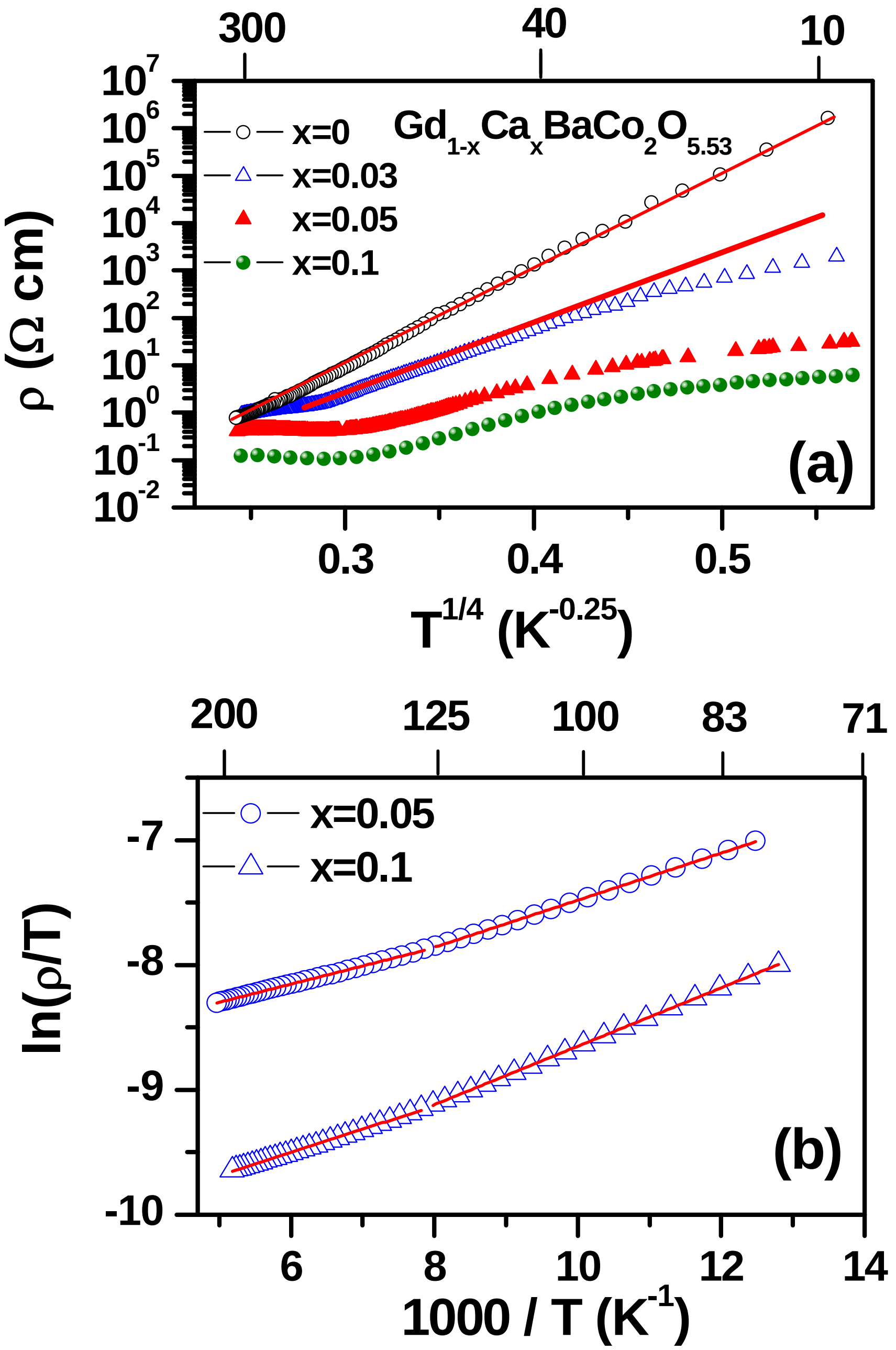

Figure 4

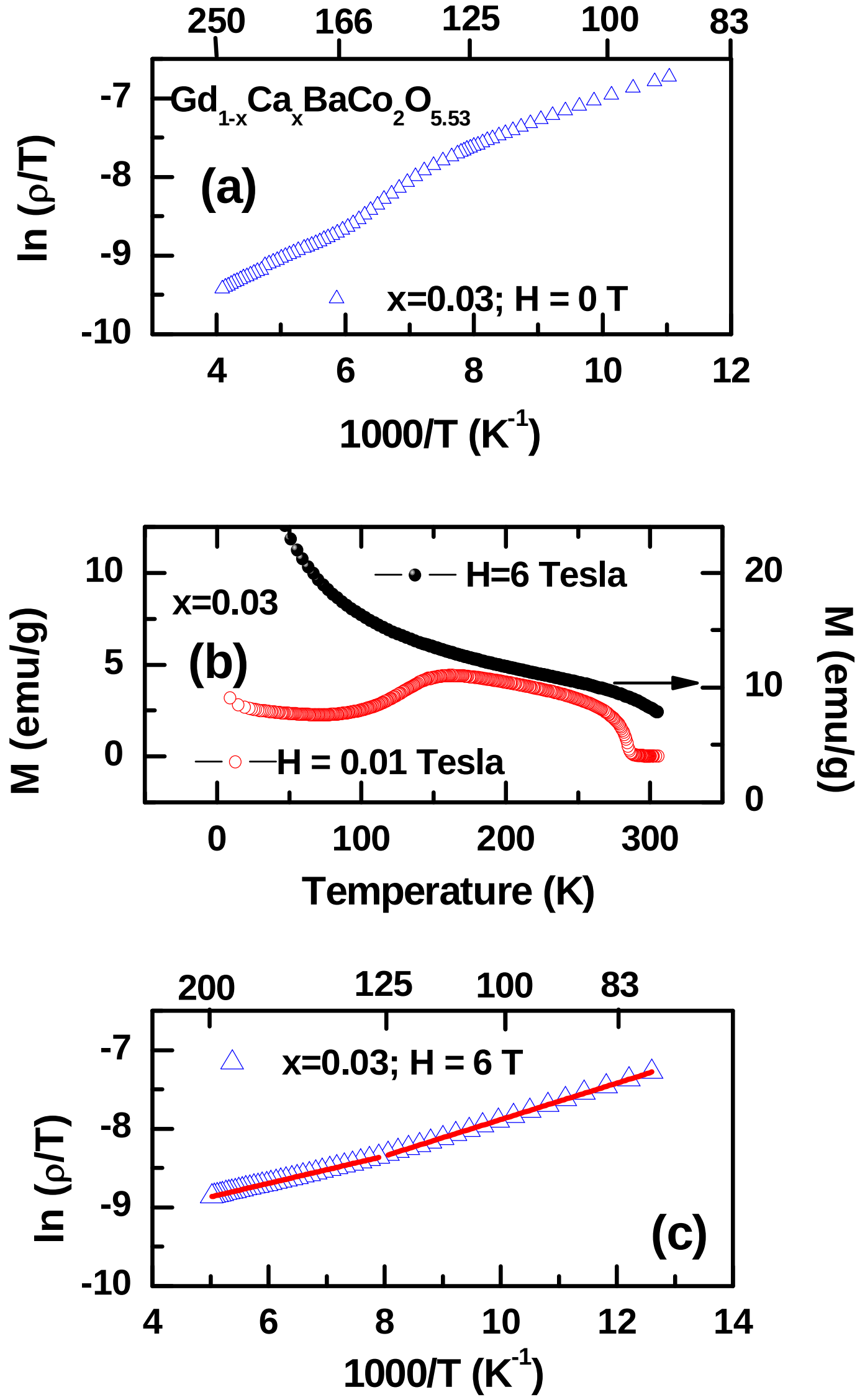

Figure 5

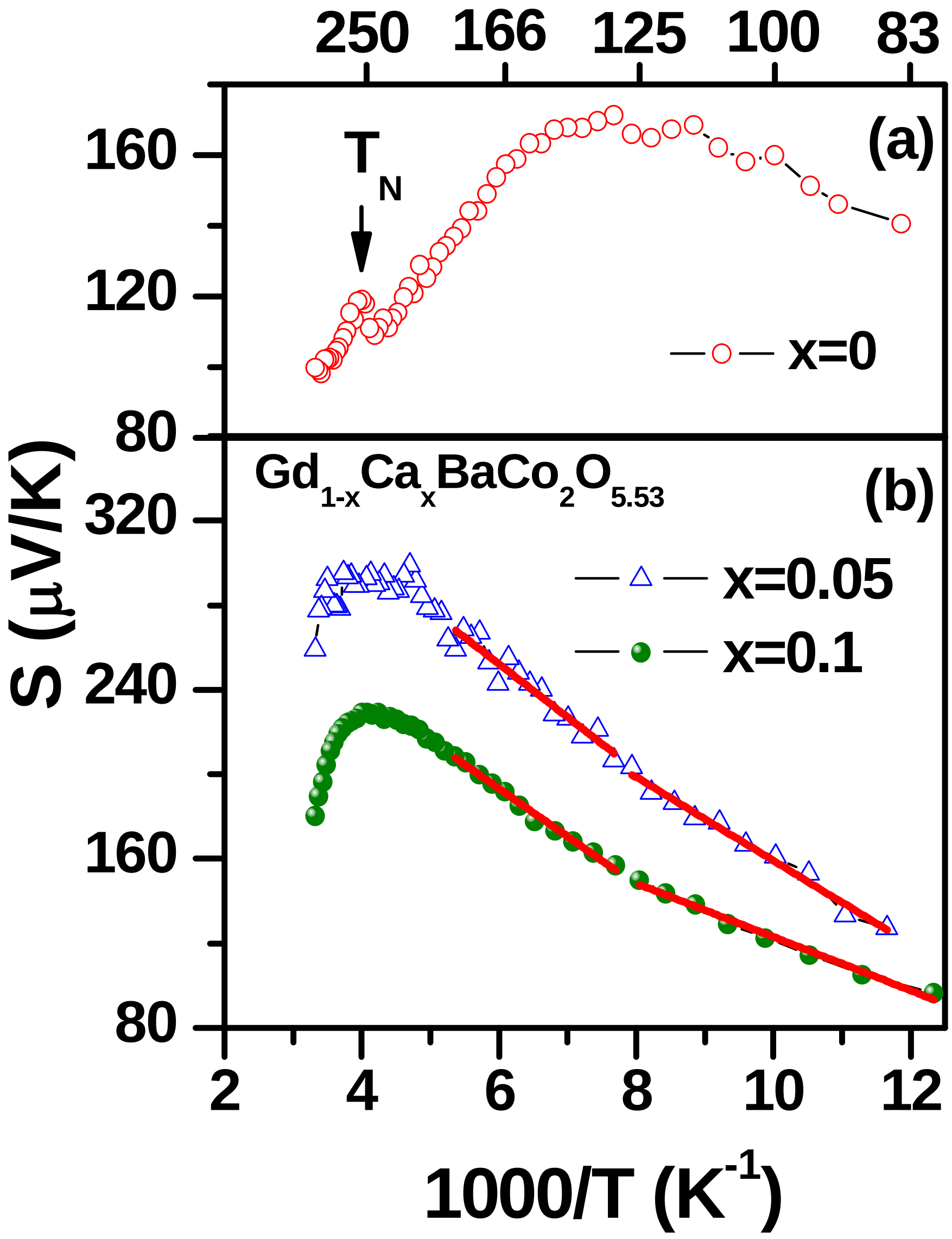

Figure 6